\documentclass[10pt,conference]{IEEEtran}
\usepackage{algorithmic}
\usepackage{graphicx}
\usepackage{textcomp}
\usepackage{xcolor}
\usepackage{multirow}
\usepackage{svg}
\usepackage{url}
\usepackage{threeparttable}
\usepackage{soul}
\usepackage{booktabs}
\usepackage{comment}
\usepackage{balance}
\usepackage[shortlabels]{enumitem}

\usepackage{tikz}

\newcommand\submittedtext{%
  \footnotesize This work has been submitted to the IEEE for possible publication. Copyright may be transferred without notice, after which this version may no longer be accessible.}

\newcommand\submittednotice{%
\begin{tikzpicture}[remember picture,overlay]
\node[anchor=south,yshift=10pt] at (current page.south) {\fbox{\parbox{\dimexpr0.65\textwidth-\fboxsep-\fboxrule\relax}{\submittedtext}}};
\end{tikzpicture}%
}

\title{Auxiliary Artifacts in Requirements Traceability: A Systematic Mapping Study}

\author{\IEEEauthorblockN{Waleed Abdeen\IEEEauthorrefmark{1},
Michael Unterkalmsteiner\IEEEauthorrefmark{1} and Krzysztof Wnuk\IEEEauthorrefmark{1}}
\IEEEauthorblockA{\IEEEauthorrefmark{1}\textit{Blekinge Institute of Technology}, 
Karlskrona, Sweden \\ 
Email: waleed.abdeen@bth.se,
michael.unterkalmsteiner@bth.se,
krzysztof.wnuk@bth.se \\
}}

\begin{document}

\maketitle

\submittednotice

\begin{abstract}
\emph{Background:} Traceability between software artifacts enhances the value of the information those artifacts contain, but only when the links themselves are reliable. Link quality is known to depend on explicit factors such as the traced artifacts and the expertise of the practitioner who judges each connection. Other factors, however, remain largely unexplored. We contend that one of these factors is the set of auxiliary artifacts---artifacts that are produced and/or used during the tracing process yet are neither the source nor target artifacts. Because such auxiliary artifacts can subtly steer how links are created and validated, they merit a literature survey to identify these artifacts and further investigate them.
\emph{Objective:} We identify and map auxiliary artifacts used in requirements tracing, which could be additional factors that affect the quality of the trace links.
\emph{Method:} We conducted a systematic mapping study on auxiliary artifacts in requirements traceability.
\emph{Results:} We found 110 studies in which auxiliary artifacts are used in requirements tracing, and identified 49 auxiliary artifacts, and 13 usage scenarios. 
\emph{Conclusion:} This study provides a systematic mapping of auxiliary artifacts in requirement tracing, including their usage, origin, type and tool support. The use of auxiliary artifacts in requirements tracing seems to be the norm, thus, these artifacts should be studied in depth to identify how they effect the quality of traced links.

\end{abstract}

\maketitle

\section{Introduction}
Requirements traceability is the ability to trace the life-cycle of requirements back to its origin (pre-requirements traceability) or a forward to downstream development artifacts (post-requirements traceability)~\cite{gotel_analysis_1994,gotel2012traceability}. Trace links between artifacts increase the value and usefulness of the information recorded in them. In requirements traceability, the most common usage scenarios for trace links are finding the origin and rationale of requirements, tracking requirement implementation state, analyzing requirements coverage in source code, and developing test cases based on requirements~\cite{bouillon2013survey}. 

Recently, there has been an increased interest in software traceability in general and requirements traceability in particular~\cite{charalampidou_empirical_2021}. However, traceability still suffers from multiple challenges when adapted in industry~\cite{ruiz_why_2023}. One of these challenges is the increased complexity of traceability practices, which results in many of the traceability techniques not being adopted or failing. Thus, it is important to study traceability techniques more in depth and understand what dependencies they have on other artifacts, as an increased dependency increases complexity. Ali et al.~\cite{ali_factors_2012} have identified three main factors that affect the quality of generated trace links between requirements and source code mainly expert opinion, requirements and source code. We argue that many of the traceability techniques use additional artifacts that are not described in literature. We refer to them as auxiliary artifacts---artifacts that are produced and/or used during the tracing process yet are neither the source nor target artifacts. In other words, an auxiliary artifact exists independently of the traced artifacts or originates as a result of traced artifact processing to aid the creation or usage of trace links. Identifying these artifacts and understanding their usage is the first step towards performing more in depth studies on how they may affect the quality of the generated trace links.

Charalampidou et al.~\cite{charalampidou_empirical_2021} have conducted a systematic mapping study (SMS) on software traceability, where they collected and mapped the traced artifacts, study goals, quality attributes, and research methods used. However, no data regarding auxiliary artifacts used in tracing were extracted from the studies. To fill this gap, we conducted this systematic mapping study.

The reminder of this paper is organized as follows. In Section~\ref{sec:method} we present our research questions and the research methodology followed. We present the results of the study, answering our research questions, in Section~\ref{sec:results}. In Section~\ref{sec:discussion} we discuss the results. Finally, we conclude the study and present the future work in Section~\ref{sec:conclusion}.

\section{Study Design}\label{sec:method}

The aim of this study is to investigate the auxiliary artifacts used in requirement traceability that are different from the source and target artifacts. To address this aim, we define the following research questions:

\begin{description}

\item [RQ1] Which auxiliary artifacts are used to trace software requirements? 

\item [RQ2] How are auxiliary artifacts used to trace software requirements? 

\item [RQ3] What is the existing tool support for auxiliary artifacts?

\item [RQ4] What artifacts are traced using auxiliary artifacts?

\end{description}

Due to this narrow scope of the research questions, we chose \emph{not} to conduct a systematic mapping study from scratch but rather to reuse an existing recent review from 2021 (Charalampidou et al.~\cite{charalampidou_empirical_2021}) and complement it by snowball sampling~\cite{wohlin_guidelines_2014}, and by reviewing publications from top conferences to identify more potentially relevant and recent studies. We follow the guidelines of Kitchenham et al.~\cite{kitchenham_guidelines_2007} for conducting and reporting SMS.

The first author conducted the SMS, performing the data collection, extraction, and reporting of the review. The second author identified the seed set and participated in pilot data extraction. Both the first and second authors analyzed the results. The third author contributed to the study design.  

\subsection{Inclusion and Exclusion Criteria}

We have defined inclusion and exclusion criteria to select primary studies, in alignment with the context and aim of our study, as follows:

\paragraph{Inclusion}

\begin{description}
    \item[I1] Traces requirements artifacts to other requirements or other software artifacts
    \item[I2] Traceability is the main topic
    \item[I3] Introduces or evaluates a requirements traceability technique 
    \item[I4] Published in a peer-reviewed venue
\end{description}
\paragraph{Exclusion}
\begin{description}

\item [E1] Not a research article (e.g., slides, book)
\item [E2] Does not investigate requirements traceability
\item [E3] Secondary Study (literature reviews)
\item [E4] Not available in Full-Text online
\item [E5] Not written in English

\end{description}

\subsection{Data Collection}
We identified research papers in the area of requirements traceability from three sources. First, we started with the data set from the systematic mapping study conducted by Charalampidou et al.~\cite{charalampidou_empirical_2021}. At the time when we started conducting this mapping study, this study was the most recent literature survey looking into software traceability research that we could find.

Second, we identified studies using snowball sampling, a study identification technique that uses a seed set to find relevant studies on a specific research topic. It is recommended to use snowball sampling to increase the number of relevant papers and reduce noise~\cite{webster_analyzing_2002,wohlin_guidelines_2014}. We used seven studies (\cite{rilling_automatic_2007,dzung_improvement_2009,kof_ontology_2010,assawamekin_ontology-based_2010,moser_requirements_2011, li_ontology-based_2013,dermeval_applications_2016}) as the seed, chosen by the second author. The seeds focused on traceability of requirements using any knowledge organization structure, one type of auxiliary artifact. We have done one iteration of sampling (i.e., one forward and one backward). The first author read the title and abstract to decide whether to include or exclude the paper. When a decision could not be made, he also read the full text.

Third, we collected studies recently published (2021-2024) in top relevant conferences: \emph{International Conference on Software Engineering} (ICSE), \emph{International Symposium on Empirical Software Engineering and Measurement} (ESEM), \emph{Requirements Engineering} (RE) and \emph{Automation Software Engineering} (ASE). Figure~\ref{fig:literature survey - study selection} illustrates the number of papers that were collected from each source.

\subsection{Data extraction}
We extracted six data elements from the included studies to answer our research questions; the data points with alignments to the research questions are presented in Table~\ref{tab:data_points}. We reused one data item (traced artifacts) extracted by Charalampidou et al.~\cite{charalampidou_empirical_2021}, and extracted the remaining five ourselves. A pilot data extraction on a set of 20 papers was done by the first and second authors. They compared the extracted data and discussed the differences to reach a common understanding of what constitutes an auxiliary artifact. We define an auxiliary artifact as \textit{an artifact that exists independently of the traced artifact or originates as a result of traced artifact processing to aid the creation or usage of trace links.}

\begin{table*}[ht]
    \caption{Data elements extracted from the included studies in the  literature survey}
    \resizebox{\textwidth}{!}{
    \centering
    \footnotesize
    \begin{tabular}{|p{0.15\textwidth}|p{0.22\textwidth}|p{0.47\textwidth}|p{0.18\textwidth}|}
    \hline
       \textbf{Research Question} & \textbf{Data element} & \textbf{Description} & \textbf{Extracted by Charalampidou et al.\cite{charalampidou_empirical_2021}} \\ 
        \hline
        \multirow{3}{0.2\textwidth}{RQ1} 
        & Existence of auxiliary artifacts 
        & whether a third artifact is used to trace the source and target artifact
        & no
        \\
        & Auxiliary artifact(s) 
        & the auxiliary artifact that was used in the traceability process
        & no
        \\
        & Auxiliary artifact's origin
        & the origin of the information used to construct the auxiliary artifact
        & no
        \\ \hline
        RQ2
        & Auxiliary artifact's usage
        & the purpose of using the auxiliary artifact in tracing
        & no
        \\ \hline
        RQ3
        & Auxiliary artifact's tool 
        & tool support for the generation or manipulation of the auxiliary artifact
        & no
        \\ \hline
        RQ4
        & Traced artifacts 
        & the source and target artifacts
        & yes
        \\
        \hline
    \end{tabular}
    }
    \label{tab:data_points}
\end{table*}

\subsection{Threats to Validity}

There are two main threats to the validity of our study: construct and internal validity.

\emph{Construct validity} concerns the design and implementation of the study. One threat is missing literature that is relevant to our topic. We mitigated this threat by method triangulation. We collected data from a recently published SMS, Charalampidou et al.~\cite{charalampidou_empirical_2021}, snowball sampling from a starting set of seven studies, and top requirements engineering conferences.

\emph{Internal validity} refers to the possible bias during data collection and analysis of the literature survey. We mitigated this threat by involving two researchers during data collection and analysis. A pilot data extraction on a set of 20 papers was done by the first and second authors. They compared the extracted data and discussed the conflicts. We found that we had a different understanding of what an auxiliary artifact is, aligned the definition, and re-extracted the data from the studies again.

\section{Results}\label{sec:results}

\begin{figure}
    \centering
    \includegraphics[width=\columnwidth]{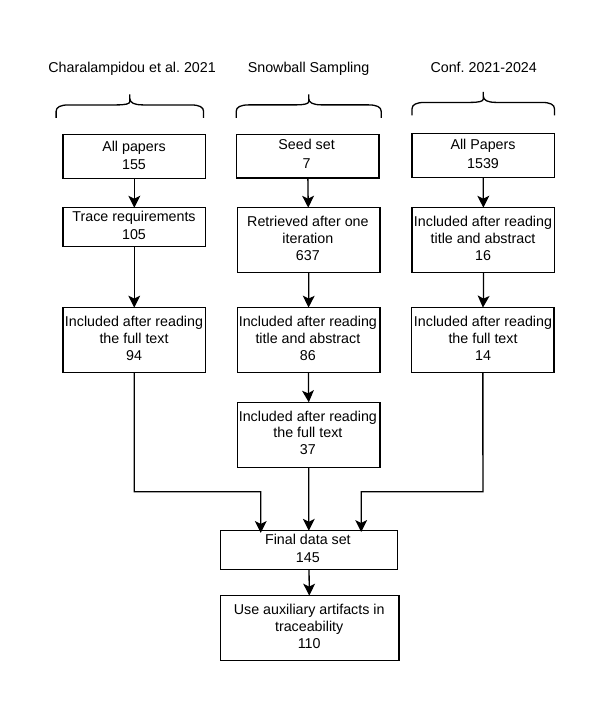}
    \caption{Literature Survey - Study Selection}
    \label{fig:literature survey - study selection}
\end{figure}

Figure~\ref{fig:literature survey - study selection} shows the results of the study selection steps. The review by Charalampidou et al.~\cite{charalampidou_empirical_2021} contained 155 studies. We selected 105 studies related to requirements traceability, 94 of which remained after reading the full text. Then, we retrieved 637 papers after one (forward and backward) snowball sampling iteration. We read the title and abstract and included 86 studies; 37 of these papers remained after reading the full text. After that, we retrieved 1539 studies published during the period 2021-2024 in top requirements related conferences and included 14 relevant studies. After consolidating the studies from all sources, we had 145 studies from which we extracted data. One-hundred and ten of these papers use auxiliary artifacts in traceability. We present the results of the study and answer our research questions in this section.

\subsection{RQ1: Auxiliary Artifacts} 

Table~\ref{tab:auxiliary_artifacts} lists the auxiliary artifacts that we extracted from the included papers. In total, we extracted 49 auxiliary artifacts used in requirements tracing to other requirements other software artifacts. The frequency in the table refers to the number of studies that use these artifacts. The top used artifacts are: \emph{terms vector} (24), a vector representation of text in n-dimensional space that can be used to measure how close these texts are in terms of similarity, \emph{terms by document matrix} (23), a two dimensional matrix that maps a document to the most frequent terms, documents with similar frequent terms deemed to be relevant, and \emph{ontology} (21), a model of concepts that represent a domain or a subject area with various relationship and properties connecting them~\cite{garshol_metadata_2004}. We classified the auxiliary artifacts on two dimensions: type and origin. The type classification is constructed bottom-up, as there was no established classification for auxiliary artifacts before this study. We identify four main types of auxiliary artifacts:

\begin{enumerate}
    \item Activity: these are artifacts that are used and/or created specifically for the tracing activity. Some examples are: terms vector~\cite{guo_tackling_2017,wang_automated_2020,saini_automated_2021}, mainly used as part of information retrieval techniques and terms by document matrix~\cite{rempel_towards_2013,diaz_using_2013,bavota_enhancing_2014,mahmoud_role_2015}, a representation of the traced artifact to ease creating trace links.
    \item Structured Knowledge: domain knowledge captured in an organized structure, which may contain domain concepts or entities. These structures are normally external to the software product, and usually describe the problem domain. E.g., ontology~\cite{cleland-huang_towards_2014,almendra_arcade_2023,mosquera_ontology-based_2023}, thesaurus~\cite{sundaram_assessing_2010,pandanaboyana_requirements_2013} and taxonomies~\cite{khan_concern_2009,unterkalmsteiner_early_2020}. 
    \item Technical: software development artifacts that are not necessarily created for tracing but are normally created or produced during the development process or during runtime. E.g., change events~\cite{mader_rule-based_2008,mader_towards_2012}, execution traces~\cite{lafet_empirical_2011,maia_impact_2013} and test logs~\cite{egyed_supporting_2005,shahid_change_2016}.
    \item Project: artifacts belonging to the project management rather than development activities. E.g., Project glossary~\cite{zou_phrasing_2006}.
\end{enumerate}

The artifact origin classification captures whether the artifact originates from one of the traced artifacts or not. E.g., terms from traced artifacts~\cite{javed_towards_2016,deshpande_requirements_2020,wang_automated_2020} originate from a text (document) that is either the source or the target of the trace link. Whereas an ontology~\cite{cleland-huang_towards_2014,almendra_arcade_2023} is external, i.e. does not originate from the traced documents.

The list of auxiliary artifacts may not be exhaustive of all artifacts used in requirements tracing, as other implicit artifacts may have been used but not clearly stated in the papers we reviewed. 

\begin{table*}[htbp]
\caption{Auxiliary Artifacts Used to Trace Requirements}
\centering
\begin{tabular}{|l|l|r|l|p{0.30\textwidth}|}
\hline

\textbf{Type} & \textbf{Auxiliary Artifact} & \textbf{Frequency} & \textbf{Origin} & \textbf{Tool Support}\\ 
\hline

\multirow{30}{0.15\textwidth}{Activity} &
    terms vector & 24 & traced artifact
    & IR-Based, Poirot, prototype, RETRO, SharpNLP, trace link evolver, TraceLab, TraceLab plugin, WordNet.NET
    \\ 
    & terms by document matrix & 23 & traced artifact 
    & ADAMS-ReTrace, blue matrix calculator, COCONUT (eclipse plugin), IR-Based, LSI, prototype, RETRO, text to matrix generator, TraceLab
    \\ 
    & training dataset & 13 & traced artifact
    & no
    \\
    & terms from traced artifact & 8 & traced artifact &
    CASE tool extension, OpenReq-DD, Poirot, TraceLab
    \\ 
    & data corpus & 6 & external
    & CLAWS4, Github
    \\ 
    & probability distribution & 4 & traced artifact
    & IR-Based
    \\ 
    & probabilistic model & 4 & traced artifact
    & prototype, TRASE, ACTS, TEAM
    \\ 
    & traceability rules & 3 & external
    & XtraQue
    \\ 
    & relational topic model & 2 & traced artifact
    & TraceLab
    \\ 
    & intermediate traces  & 2 &  traced artifact
    & Neo4j
    \\
    & action units & 1 & traced artifact 
    & no
    \\ 
    & document embedding vector & 1 & traced artifact
    & no
    \\
    & document topic vector & 1 & traced artifact 
    & no
    \\
    & eye tracking info & 1 & external
    & Online meeting tool
    \\
    & flexible-pla metamodel & 1 & traced artifact 
    & FPLA (eclipse plugin)
    \\
    & inference rules & 1 & external
    & no
    \\ 
    & keywords & 1 & external
    & FPLA (eclipse plugin)prototype
    \\ 
    & linkage rules & 1 & external 
    & no
    \\ 
    & linguistic model & 1 & external
    & no
    \\ 
    & phrases from traced artifact & 1 & traced artifact
    & prototype
    \\ 
    & program data table & 1 & traced artifact
    & eclipse plugin
    \\ 
    & requirements metamodel & 1 & traced artifact 
    & no
    \\
    & semantic relation graph & 1 & traced artifact 
    & no
    \\ 
    & sequential patterns & 1 & traced artifact
    & no
    \\ 
    & transformation rules & 1 & external
    & no
    \\ 
    & tags & 1 & external 
    & eclipse plugin
    \\     
    & validation rules & 1 & external 
    & no
    \\ 
    & trace ranking & 1 & traced artifact
    & prototype
    \\
    & trace model & 1 & traced artifact
    & no
    \\ 
    & metamodel & 1 & external &
    DesignSpace
    \\
    \hline
\multirow{6}{0.15\textwidth}{Structured Knowledge} 
    & ontology & 21 & traced artifact, external 
    & MUPRET, OntoGazetteer, ONTrace, Protégé, Semantic mission profile aware design platform, ARCADE, OntoTrace
    \\ 
    & thesaurus & 9 & external &
    thesaurus builder
    \\ 
    & taxonomy & 4 & traced artifact, external
    & RETH
    \\ 
    & domain model & 2 & traced artifact, external
    & domain model extraction too
    \\ 
    & wikipedia & 1 & external
    & no
    \\
    & concept map & 1 & traced artifact 
    & OntoLancs
    \\ 
    \hline
\multirow{10}{0.15\textwidth}{Technical} &
    change events & 3 & traced artifact
    & event generator plugin
    \\ 
    & execution traces & 3 & traced artifact
    & Aspectj, prototype
    \\ 
    & xml document & 3 & traced artifact
    & prototype, XtraQue
    \\ 
    & test logs & 2 & traced artifact
    & compiler
    \\ 
    & call graph & 2 & traced artifact 
    & multi-threaded tracer, exploring traces
    \\ 
    & feature metamodel & 2 & traced artifact 
    & FPLA (eclipse plugin)
    \\ 
    & block level requirements & 1 & traced artifact 
    & SafeSlice
    \\ 
    & early aspects & 1 & traced artifact
    & EA-Miner
    \\ 
    & code aspects & 1 & traced artifact
    & FINT
    \\
    & quality types & 1 & external 
    & no
    \\ 
    \hline
\multirow{3}{0.2\textwidth}{Project} & 
    project glossary & 2 & external
    & no
    \\ 
    & work items & 1 & traced artifact 
    & no
    \\ 
    & process description & 1 & external 
    & no
    \\ 
\hline
\end{tabular}
\label{tab:auxiliary_artifacts}
\end{table*}

\subsection{RQ2: Auxiliary Artifacts Usage}\label{sec:usage}

\begin{figure}
    \centering
    \includegraphics[width=1\linewidth]{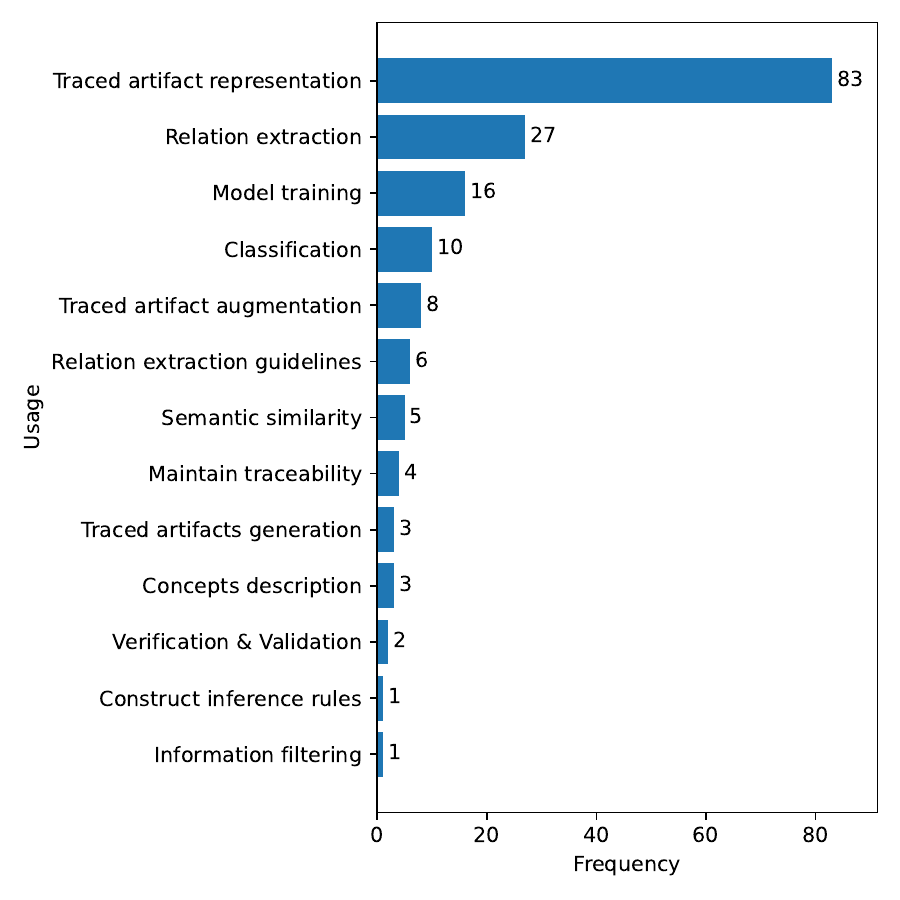}
    \caption{Auxiliary artifacts usage}
    \label{fig:aa_usage}
\end{figure}

Figure~\ref{fig:aa_usage} depicts the auxiliary artifacts usage in requirements traceability and their frequency. In general, an artifact could have one or more usages. If an artifact had more than one usage in a study, we extracted the main usage. For example, an auxiliary artifact used in the Jenson-Shannon model~\cite{lin_poirot_2006} is the probabilistic distribution of the words in documents, which has two usages: traced artifact representation and relation extraction. We reported traced artifact representation as it is the main usage. Below, we describe each usage.

\begin{enumerate}
    \item \emph{Traced artifact representation}: a representation of the traced artifacts in a form that eases processing of information to infer traces, e.g., terms vector in information retrieval (IR) methods.
    \item \emph{Relation extraction}: the extrapolation of the type and degree of relevance between two objects from the traced artifacts or from the traced artifact and the auxiliary artifact. In this usage, the semantic and structural information of the auxiliary artifact is used.
    \item \emph{Model training}: the training of machine learning models to predict trace links, e.g., a classifier or neural networks.
    \item \emph{Classification}: the use of an auxiliary artifact to classify the traced artifacts. A trace link is established between artifacts that have the same class assigned to them.
    \item \emph{Traced artifact augmentation}: the introduction of relevant terms in the traced artifact to improve the probability of finding a corresponding artifact.
    \item \emph{Relation extraction guidelines}: a set of guidelines that are used to extract relations between two objects.
    \item \emph{Semantic similarity}: regards the calculation of the similarity degree between two terms, normally using cosine similarity. 
    \item \emph{Maintain traceability}: the process of adding, removing, and modifying trace links to keep them updated.
    \item \emph{Traced artifacts generation}:  traceability is created as a side effect of the artifact generation process.
    \item \emph{Concepts description}: finding descriptions and synonyms of domain concepts that are presented in the traced artifacts or another auxiliary artifact (e.g., ontology) to improve the performance of term matching between artifacts that contain different semantics.
    \item \emph{Verification \& Validation}: to ensure the correctness of the traceability process and the created trace links.
    \emph{Construct inference rules}: the creation of rules (guidelines) to find elements of the traced artifacts that are consistent (equal, synonyms, composite, or ancestor-descended).
    \item \emph{Information filtering}: find the most relevant information by using attributes to aid the generation of trace links.
\end{enumerate}

Auxiliary artifacts in requirements engineering have different usages, many of these usages, e.g., traced artifact representation, relation extraction and model training concern the automation of traceability.

\subsection{RQ3: Tool Support for Auxiliary Artifacts}

We gathered information about the tools used to create or modify the auxiliary artifact. Among the 49 artifacts, 31 were associated with some form of tool support, while 18 were not. These tools are either traceability tools or specialized tools in managing the auxiliary artifacts. A traceability tool generates and uses the auxiliary artifact within the same tool to create traces, such as ADAMS Re-Trace~\cite{de_lucia_adams_2005,bavota_enhancing_2014}, which implements IR techniques like latent semantic indexing (LSI) and generates terms by document. A specific tool is designed solely to create the auxiliary artifact, like the blue matrix calculator~\cite{mahmoud_role_2015}, which generates terms by document matrices for LSI, which are then used to create traces manually or with another tool. The tools are presented in Table~\ref{tab:auxiliary_artifacts}.

\subsection{RQ4: Traced Artifacts}

In Table~\ref{tab:ta}, we list the artifacts traced using auxiliary artifacts and the count of the primary studies where we found them. We did not have predetermined groups, instead we grouped these artifacts using four process models: Rational Unified Process (RUP), Open Unified Process (OpenUP), 830 IEEE Standard for Software Requirements Specification, and Scrum. These process models were also used by Charalampidou et al.~\cite{charalampidou_empirical_2021} to group the classified artifacts. Originally, we identified 74 artifacts which we grouped into 22 groups, listed in Table~\ref{tab:ta}.

\begin{table}[!ht]
    \centering
    \caption{Artifacts traced using Auxiliary Artifacts}
    \begin{tabular}{|l|l|}
    \hline
        \textbf{Traced Artifact} & \textbf{Count} \\ \hline
        Requirements & 87 \\ 
        UML design diagrams & 39 \\ 
        Code classes & 29 \\ 
        Use cases & 28 \\ 
        Source code & 24 \\ 
        Test cases/test documents & 21 \\ 
        Design (general, models) & 16 \\ 
        Methods/functions/operations/package/module & 15 \\ 
        Features/functional requirements & 14 \\ 
        Design artifacts (general, documents, parameters, specs) & 6 \\ 
        Documentation & 5 \\ 
        Code objects/attributes/class descriptions & 5 \\ 
        Issues/Bug reports & 4 \\ 
        Query/emails & 4 \\ 
        Architectural artifacts/elements & 3 \\ 
        Regulations/regulatory code & 2 \\ 
        Commits & 2 \\ 
        Scenarios & 2 \\ 
        Architectural diagrams & 2 \\ 
        Runtime artifacts & 1 \\ 
        Domain models & 1 \\ \hline
    \end{tabular}
    \label{tab:ta}
\end{table}

\section{Discussion}\label{sec:discussion}

In this section, we discuss the results of the study presented in Section~\ref{sec:results}.

\subsection{Importance of Auxiliary Artifacts}
Our systematic mapping study indicates that auxiliary artifacts are common in requirements tracing. 
They fulfill varied roles and may either originate from the artifacts being traced or exist independently of them (external). The usages summarized in Section~\ref{sec:usage} indicate that these auxiliary artifacts are key enablers of trace-link automation. Yet, greater reliance on them may also increase the complexity of traceability practices~\cite{ruiz_why_2023}. 

\subsection{Effect on Traceability Maintenance}
Traceability maintenance is a significant cost driver that can hinder the adoption of traceability links~\cite{ruiz_why_2023} and distort traceability measurements~\cite{maro_software_2018}. It is therefore essential to consider every factor that demands attention when creating and maintaining trace links; otherwise, costs escalate and the return on investment becomes elusive. Auxiliary artifacts are one such factor: they influence the effort of establishing and preserving traceability and, if overlooked, can render a trace solution ineffective. For instance, in automated trace‑retrieval approaches~\cite{guo_tackling_2017}, a \emph{data corpus} is used to expand the query and mitigate the term‑mismatch problem between the query and the target requirements. However, as the software solution evolves, this corpus can quickly become outdated, leading to inaccurate trace links and, consequently, degraded traceability maintenance.

\subsection{Traceability without Auxiliary Artifacts}
Out of the 145 studies that we identified, 35 did not mention using any auxiliary artifacts for tracing. Further analysis of these studies revealed two observations. Firstly, trace links were created manually either by practitioners~\cite{mader_assessing_2012} or students~\cite{egyed_effort_2010} by analyzing the traced artifacts (e.g., requirements and source code). In some cases, tools were provided to facilitate the creation~\cite{ji_maintaining_2015} and visualization of the links~\cite{jaber_study_2013}. Secondly, there was a lack of information on how trace links were generated in some of these studies. Although they may have used auxiliary artifacts for tracing, it is not clear which artifacts were used. For example, Cleland-Huang~\cite{cleland-huang_breaking_2012} used ADAMS, in addition to other IR-based traceability tools, to generate trace links. ADAMS generates and uses multiple auxiliary artifacts~\cite{de_lucia_adams_2005,bavota_enhancing_2014}, depending on the IR method used, to create trace links. Furthermore, a lack of information could be due to these studies focusing on other parts of the traceability process besides trace links creation, e.g., trace links validation~\cite{ghabi_code_2012}.

\subsection{Traced Artifacts}
The majority of studies focused on tracing \emph{requirements} with no specific type or style mentioned (87), or a specific type of requirements, \emph{use cases} (28), or \emph{features} (14). These artifacts were traced to artifacts found in different stages of the development, e.g., \emph{UML diagrams} in software design, \emph{source code} and \emph{classes} in development, and \emph{test cases} in testing. However, tracing to artifacts that are produced outside the development process is the least studied. We found only six instances of artifacts outside the development process where requirements are traced to:\emph{domain models} and \emph{query/emails} and \emph{regulations}.

\section{Conclusion and Future Work}\label{sec:conclusion}
We have conducted a systematic mapping study on auxiliary artifacts for requirements traceability. To the best of our knowledge, this is the first study that identifies explicitly the auxiliary artifacts that are used in tracing.Our findings for RQ1 indicate that auxiliary artifacts play a crucial role in requirements traceability and should be carefully managed. Moreover, the increased reliance on auxiliary artifacts introduces additional complexity in traceability, particularly in terms of information processing and management. Regarding the use of auxiliary artifacts (RQ2), we found that they primarily serve to provide supplementary information to trace link creators/maintainers, or automated approaches. This suggests that the information contained in primary artifacts alone is often insufficient to effectively uncover traceability relationships. Overall, many of the identified auxiliary artifacts have one or more associated tool that supports its management (RQ3). A significant number of the development~\emph{activity} related artifacts and \emph{project} related artifacts lack tool support in relation to traceability. As for RQ4, auxiliary artifacts are used in tracing requirements to multiple types of artifacts produced as part of different stages of the development; those that are least traced are architectural diagrams and external documents (e.g., regulations and standards). 

Future work concerns investigating how much complexity these artifacts are adding to traceability practices how they are maintained. Moreover, software engineering researchers should think about how these artifacts could affect the quality of traces and identify a quality criteria for them.

\section{Data Availability}\label{sec:data availability}
We provide additional material online\footnote{https://figshare.com/s/b36c0db63ddb8ebb80cc} containing a list of the included papers and the extracted data. We will generate a DOI for the materials upon acceptance of the paper.

\balance
\bibliographystyle{IEEETrans}
\bibliography{IEEEabrv,refrences}


\end{document}